 \documentclass[prb,twocolumn,showpacs, ,amssymb,superscriptaddress]{revtex4}
 \usepackage{graphicx}
\usepackage{dcolumn}
\usepackage[sort&compress]{natbib}
\usepackage{subfigure}
\usepackage{ifpdf}
\usepackage{bm}
\ifpdf
\usepackage[pdftex,
        colorlinks=true,
        pdftitle={},
        pdfauthor={S. Salem-Sugui Jr.},
        pdfsubject={fluctuation in iron-pnictides},
        pdfkeywords={IF, UFRJ, Rio de Janeiro Brazil},
        baseurl={http://www.if.ufrj.br},
        pdfpagemode=UseNone,
        bookmarksopen=true
        pdfoutput=1
        ]{hyperref}
\fi
\begin{document}
\title{Flux-creep in the second magnetization peak of BaFe$_{1.9}$Ni$_{0.1}$As$_2$ superconductor. }
\author{S. Salem-Sugui Jr.}
\affiliation{Instituto de Fisica, Universidade Federal do Rio de Janeiro,
21941-972 Rio de Janeiro, RJ, Brazil}
\author{L. Ghivelder} 
\affiliation{Instituto de Fisica, Universidade Federal do Rio de Janeiro,
21941-972 Rio de Janeiro, RJ, Brazil}
\author{A. D. Alvarenga}
\affiliation{Instituto Nacional de Metrologia Normaliza\c{c}\~ao e
Qualidade Industrial, 25250-020 Duque de Caxias, RJ, Brazil.}
\author{L.F. Cohen}
\affiliation{The Blackett Laboratory, Physics Department, Imperial College London, London SW7 2AZ, United Kingdom}
\author{ Huiqian Luo}
\affiliation{Beijing National Laboratory for Condensed Matter Physics, Institute of Physics, Chinese Academy of Sciences,  Beijing, 100190, P. R. China.}
\author{Xingye Lu}
\affiliation{Beijing National Laboratory for Condensed Matter Physics, Institute of Physics, Chinese Academy of Sciences,  Beijing, 100190, P. R. China.}
\date{\today}
\begin{abstract}
Flux-creep data was obtained for fields along the second magnetization peak  observed in $M(H)$ curves of BaFe$_{1.9}$Ni$_{0.1}$As$_2$ for  $H$$\parallel$$c$-axis. $H$$\parallel$$ab$-planes and $H$ forming a 45$^o$ angle with $ab$-planes. The $M$-$H$ loops from the different field directions can be collapsed onto a universal curve with a scaling factor equivalent to the superconducting anisotropy, showing that the pinning is three dimensional, although with remarkable  differences in the vortex-dynamics as a function of field orientation. The resulting relaxation rate, $R$,  when plotted as a function of field and temperature does not show any specific feature in the vicinity of the second magnetization peak field $Hp$, the relaxation shows a maximum at a field $H_2$ well above $Hp$ for $H$$\parallel$$c$ and a minimum at $H^*$ for fields well below $Hp$ for $H$$\parallel$$ab$ and $H$-45$^o$-$ab$. Isofield plots of the scaled activation energy obtained from flux-creep data at several different temperatures also do not show any evidence of a change in the pinning mechanism as the peak field is crossed. The $Hp$ lines in the resulting phase diagrams do not appear to be consistent with a description-terms of a collective-plastic pinning crossover. 
\end{abstract} 
\maketitle 
The study of vortex-dynamics in the novel pnictide superconductors\cite{1} has attracted increasing attention, due to the considerable high-$T_c$ of these compounds, when compared to the conventional superconductors, and because of similarities with the high-$T_c$ cuprates superconductors \cite{norman}. Most of the pnictides systems exhibit the second magnetization peak, or fish-tail, in isothermic magnetization curves, as well as large flux-creep, allowing the study of different regions of the vortex-phase diagram in certain detail\cite{2,4,5,6,7,8,9,9b}. The second magnetization peak is associated with a maximum in the critical current when measured as a function of field at a fixed temperature. This phenomena is not yet completely understood in pnictides\cite{4}. Although pnictides have similarities with the cuprates, such as the layered structure and antiferromagnetism of the precursor non-superconducting system \cite{norman}, it is well established that superconductivity in pnictides has a multi-band characther\cite{tesa} for which it is predicted the existence of non-Abrisosov vortices\cite{10,11} which might lead to new effects in the vortex matter. In this work we study the vortex dynamics in a BaFe$_{1.9}$Ni$_{0.1}$As$_2$ superconductor single crystal, by means of isofield magnetization $M(H)$ curves and magnetic relaxation $M(t)$ curves.

Magnetization data were obtained by using commercial magnetometers: a 5T MPMS based on a superconducting quantum interference device (SQUID) was used for most of measurements with $H$$\parallel$$ab$-planes ; and a 9T PPMS was used for the other measurements including all data with $H$$\parallel$$c$-axis and and $H$-45$^o$-$ab$-planes. The measurements were made after lowering the sample temperature from above $T_c$ in zero applied magnetic field (ZFC-procedure). The studied sample, is a high-quality single crystal of BaFe$_{1.9}$Ni$_{0.1}$As$_2$ with transition temperature $T_c$ = 20 K, transition width $\Delta$$T_c$=0.3 K, mass = 120 mg and dimensions 0.5x1.6x0.02 cm. Details of the sample preparation can be found in Ref. \onlinecite{luo}. For $H$$\parallel$$c$-axis and $H$-45$^o$-$ab$ geometries, we carefully broke the sample and used a 43.1 mg piece with dimensions ~0.55x0.5x0.02 cm. For  $H$$\parallel$$ab$-planes, the sample was attached to a hard-plastic slab perfectly inserted along the entire length of the straw tube used in the measurement systems, assuring an almost perfect alignment of field with the $ab$-planes. 

\begin{figure}[t]
\includegraphics[width=\linewidth]{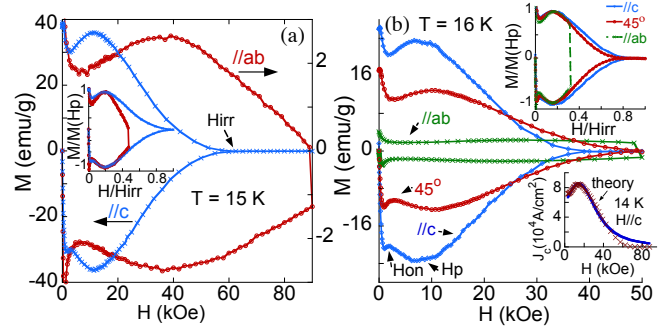}
\caption{a) double plot of $M(H)$ at 15 K for $H$$\parallel$$c$ and $H$$\parallel$$ab$.; b) $M(H)$ curves at 16 K for the three geometries. Insets: a) and upper b) shows a reduced plot of the same correspondent $M(H)$ curves.; lower b), $J_c(H)$ at 14 K for $H$$\parallel$$c$.}
 \label{fig1}
\end{figure}
Figure 1a shows a double plot of $M(H)$ curves obtained at 15 K for $H$$\parallel$$c$ and $H$$\parallel$$ab$ where it is possible to observe the differences in the shape of the fish-tail between both field directions, which as shown below, are due to differences in the vortex dynamics. Figure 1b shows $M(H)$ curves at 16 K for all field directions, all exhibiting the second magnetization peak, where $Hon$ represents the onset of the second magnetization peak with its maximum value occurring at $Hp$. The inset of Fig. 1a and the upper inset of Fig. 1b show a reduced plot of the curves appearing in the respective main figures, where $M$ is divided by the respective value at $Hp$ and $H$ by the respective irreversible field $Hirr$. These inset figures show that $Hirr$ and $Hp$, for the different geometries, can be collapsed onto a universal $M$-$H$ curve. The lower inset of Fig. 1b shows a curve of the critical current $J_c(H)$ at $T$ = 14 K for $H$$\parallel$$c$, which was obtained from the correspondent $M(H)$ curve by using the Bean Model \cite{bean}. The almost perfect symmetry of the $M(H$) curves of Fig. 1 with respect to the $x$-axis  evidence that bulk pinning is dominant. Also, the equilibrium magnetization, $Meq$, defined as the average value of $M$ on both branches of a $M(H)$ curve, is very small, so we may use values of $M$ instead ($M-Meq$) in the analysis that follows. Near $T_c$ for temperatures above 19 K, the second magnetization is no longer observed. 

The vortex dynamics study was performed by collecting isofield magnetic relaxation data, $M(t)$ curves, along several isothermic $M(H)$ curves, for fields lying below and above the second magnetization peak, for the three geometries. We also obtained isofield magnetic relaxation curves as a function of temperature. Magnetic relaxation data were collect for 2 hrs when obtained in both branches of $M(H)$ curves and for 3.5 hrs when only in the lower branch. We also measured long time magnetic relaxation for approximately 12 hrs in different regions of the $M(H)$ curve for $H$$\parallel$$c$-axis. All $M(t)$ vs. log$(t)$ curves show strictly linear behavior, starting above a transient time $\tau_0$ $\approx$ 3-4 min. Such a large transient time was observed before in BaFe$_{1.82}$Ni$_{0.18}$As$_2$ with $T_c$=8 K, and seems to be intrinsic of the material \cite{8}. The linear behavior with log$(t)$  was also observed for the 12 hrs relaxation curves. This fact suggests that it is more appropriate to analyse the data by using the relaxation rate $R$ = $dM/dln(t)$ as defined in Ref. \onlinecite{beasley}. As shown in Ref. \onlinecite{beasley} one may obtain information on the apparent activation energy, but this quantity may not have any physical meaning in our data, since the magnetization $M_0$ at the time $t$ = 0, above which logarithmic relaxation should start, is not well defined, due to the $\approx$ 4 min long transient region (see the inset of Fig. 2c below).

\begin{figure}[t]
\includegraphics[width=\linewidth]{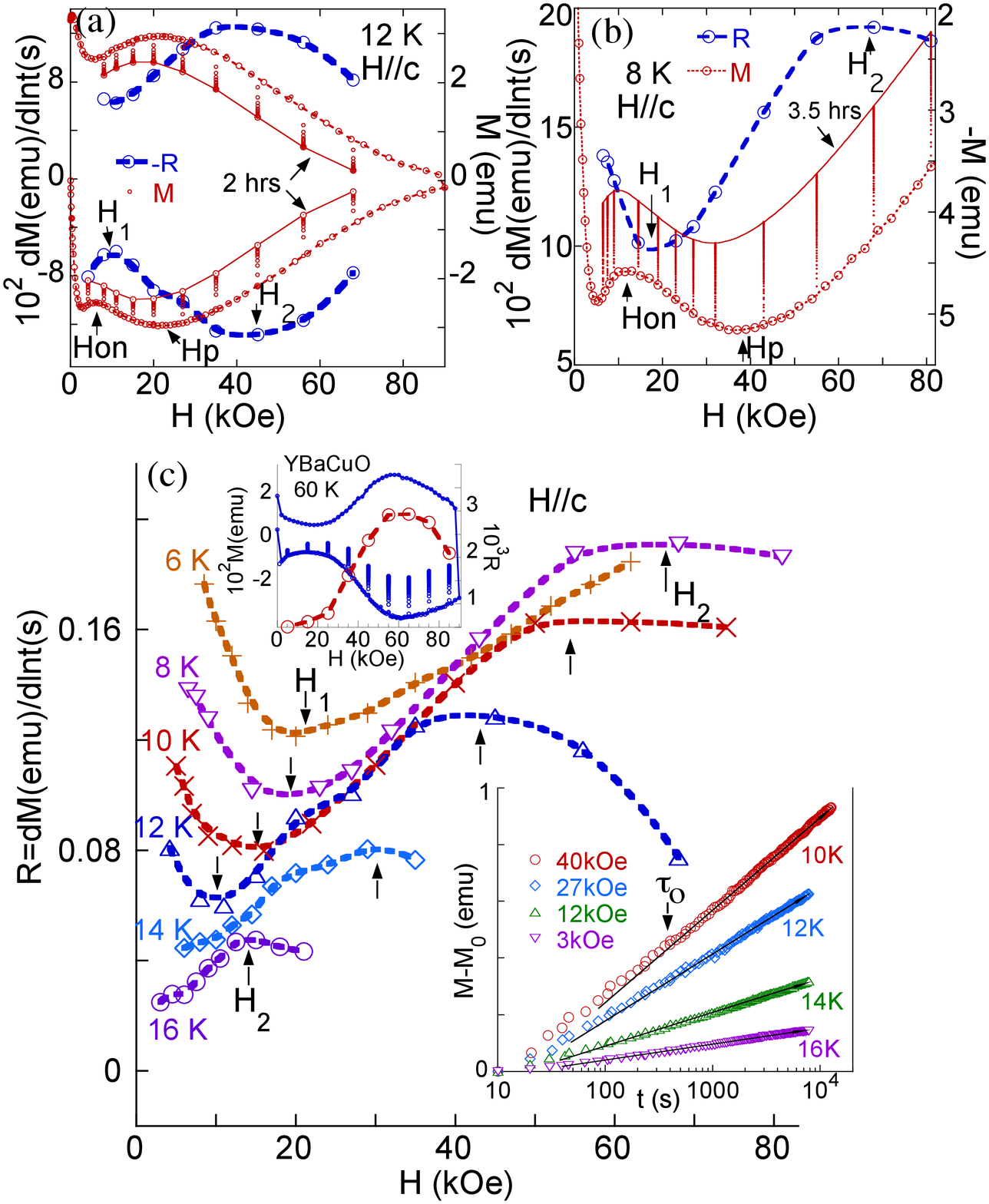}
\caption{$H$$\parallel$$c$-axis: Double plots of $-R~vs.~H$ and $M(t)~vs.~H$ at  a) $T$ = 12 K; b) $T$ = 8 K; c) $R~vs.~H$ for the lower branch of $M(t,H)$ curves. Upper inset: Double plot of $R~vs.~H$ and $M(t,H)$ for YBaCuO at $T$ = 60 K. The lower inset shows selected magnetic relaxation data.}
 \label{fig2}
\end{figure}
Figure 2 shows results of the analysis of $M(t)$ data for $H$$\parallel$$c$-axis. The logarithmic relaxation of the magnetization, is exemplified by the selected curves plotted in the inset of Fig. 2c.
Figure 2a shows a double plot of $R(H)~vs.~H$ and the corresponding $M(H)$ curves where the 2hrs relaxation data are also plotted. In Fig. 2a, we plot  $R$ as a function of increasing and decreasing field and we define $H_1$ as the minimum in $R$ ($R$ increases above $H_1$) and $H_2$ a maximum ($R$ decreases above $H_2$). It is interesting to note that the resulting $R(H)$ curve resembles the correspondent $M(H)$ curve after a shift to the right. The same trend as shown in Fig. 2a was observed for $T$ = 14 K and 16 K data. To explore whether the field $H_1$ is related to $Hon$, we took 3.5 hrs relaxation data in the lower branch of $M(H)$ curves at $T$ = 6, 8, and 10 K including more data for fields below $Hon$. Figure 2b shows the results for $T$ = 8 K with a double plot of $R~vs.~H$ and the correspondent $M(t,H)~vs.~H$, evidencing that the field $H_1$ is indeed related to the field $Ho$n, but $H_2$ is about 30 kOe above $Hp$ and these two fields are not likely to be related. It is clear from Figs. 2a and 2b, which are representative data from all experiments that there is no apparent change in the relaxation rate as the peak field $Hp$ is crossed. Figure 2c shows the results of $R~vs.~H$ for relaxation data obtained on the lower branch of all $M(H)$ curves, where the arrows pointing up and down show the approximately positions of $H_2$ and $H_1$ respectively.  For the sake of comparison, the upper inset of Fig. 2c shows a double plot, as in Fig. 2b, for a YBaCuO sample ($T_c$ $\approx$ 92 K), with relaxation data obtained during 60 min ($H$$\parallel$$c$-axis) at the lower branch of an $M(H)$ curve at $T$ = 60 K. It is clear the perfect matching between $Hp$ and the field position of the maximum in $R$, which in the case of YBaCuO \cite{abulafia}, represents a pinning crossover taking place as $Hp$ is crossed. A direct comparison of the upper inset of Fig. 2c with Fig. 2a or 2b, suggests that the second magnetization peak in these two systems arise from different mechanisms.

\begin{figure}[t]
\includegraphics[width=\linewidth]{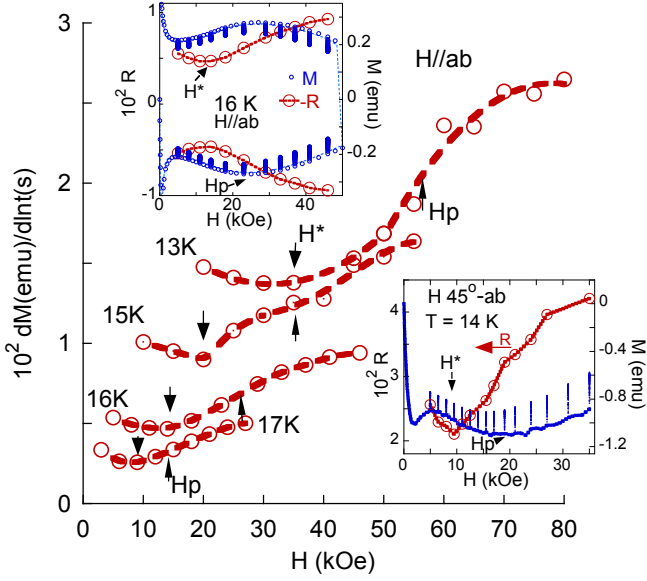}
\caption{$R~vs.~H$ for the lower branch of $M(t,H)$ curves for $H$$\parallel$$ab$-planes. Insets: Double plots of $R~vs.~H$ and $M(t,H)$; upper, at $T$ = 16 K with $H$$\parallel$$ab$; lower, at $T$ = 14 K with $H$45$^o$$ab$: }
 \label{fig3}
\end{figure}
Figure 3 shows  the results for $H$$\parallel$$ab$-planes. In that case, $R$ decreases as field increases above $Hon$, reaching a minimum at a field denominated $H^*$, which lies well below $Hp$. For each curve of Fig. 3, arrows pointing up shows the position of $Hp$ and pointing down of $H^*$. A similar result was previously observed for an overdoped BaFe$_{1.82}$Ni$_{0.18}$As$_2$ with $T_c$ = 8 K for $H$$\parallel$$c$-axis and $H$$\parallel$$ab$-planes\cite{8}. The upper inset of Fig. 3 shows a double plot of $R~vs.~H$ and the correspondent $M(t,H)~vs.~H$ curve. To check whether this change in $R$ only occur when $H$$\parallel$$ab$-planes, we measure the sample for an intermediate geometry, with $H$ forming a 45$^o$ with $ab$-planes. The results for $T$ = 14 K is shown in the lower inset of Fig. 3 (the same trend is observed for $T$ = 16 K), showing that vortex dynamics for this geometry is similar to that observed for $H$$\parallel$$ab$-planes.  
\begin{figure}[t]
\includegraphics[width=\linewidth]{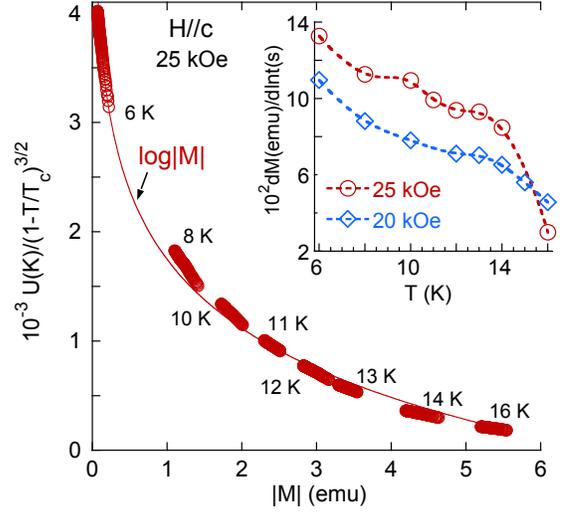}
\caption{$U(M,T)/(1-T/Tc)^{3/2}~vs.~ |M|$ for $H$ = 25 kOe, $H$$\parallel$$c$-axis. The solid line represents a $log|M|$ curve. Inset: $R~vs.~T$ for $H$ = 20 and 25 kOe, $H$$\parallel$$c$-axis.}
 \label{fig4}
\end{figure}

To check whether a pinning crossover, or a vortex phase transition, would perhaps become evident near $Hp$ using a different approach, we obtained a few isofield magnetic relaxation data as a function of temperature for  $H$$\parallel$$c$ and  $H$$\parallel$$ab$. Now, as in Ref.\onlinecite{maley2}, we calculate the pinning activation energy for a set of isofield data $M(t,T)$ using the expression, $U(M)$=-$T$ln($dM/dt$)+$CT$ where $C$=ln($Bwa/\pi d$) is a constant, where $B\approx H$ is the magnetic induction, $\omega$ is an attempt frequency, $a$ is the hop distance and $d$ is the sample size.  It is believed\cite{maley2} that the isofield $U(M,T)/g(T/T_c)$ (where $g(T/T_c)$ is an appropriated scaling function of $U$\cite{maley2}) should be a smooth function of $|M|$ within a temperature region with same pinning mechanism.  Figure 4 shows the results of $U(M,T)/g(T/T_c)$ plotted against $|M|$ for $H$$\parallel$$c$, as obtained for $H$ = 25 kOe data with $C$ = 10 and $g(T/T_c)$ = $(1-T/T_c)^{3/2}$. Similar values of the constant $C$ were found for cuprates superconductors\cite{maley2} and for pnictides\cite{4,8}. The almost perfect log$|M|$ fit linking all the data in Fig. 4 suggests the existence of only one pinning mechanism over the entire $\Delta T$ range, for which $Hp$ is located near 11 K.  The same trend was observed for $U(M,T)/(1-T/T_c)^{3/2}~vs.~ |M|$ obtained for $H$$\parallel$$ab$ data with $H$ = 15 kOe (not shown). The inset of Fig. 4a shows plots of $R~ vs.~ T$ as obtained from isofield data with $H$ = 20 and 25 kOe for $H$$\parallel$$c$, which do not show any visible effect near $Hp$ located at ~12 K and ~11 K in each respective curve. One would expect some feature in the plot of $R~vs.~T$ as $Hp$ is crossed, either for a pinning crossover or for a vortex-lattice phase transition. The same behavior was observed on similar plots for $H$$\parallel$$ab$-planes with $H$ = 15 and 20 kOe. 

\begin{figure}[t]
\includegraphics[width=\linewidth]{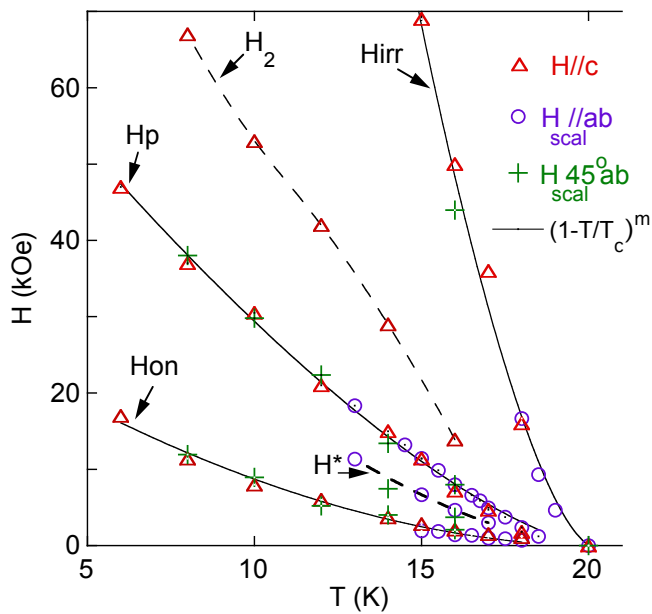}
\caption{Phase diagram for the three geometries after scale the Y-axis for $H$$\parallel$$ab$ and $H$45$^o$$ab$. Solid lines represents a fitting of the data. Dashed lines are only a guide to the eyes. }
 \label{fig5}
\end{figure}
Figure 5 shows the resulting phase diagram where characteristic fields of the three geometries are plotted against temperature after being divided by $\sqrt {(sin(\theta ))^2+(1/3)(cos(\theta ))^2}$ where $\theta$ is the angle between $H$ and the $ab$-plane, and the factor 3$\approx$$Hp$($H$$\parallel$$ab$)/$Hp$($H$$\parallel$$c$)$\approx$$Hirr$($H$$\parallel$$ab$)/$Hirr$($H$$\parallel$$c$) is of the order of the system anisotropy.\cite{abc} For clarity reasons we did not plot values of $H_1$ which lie close to $Hon$. The collapse of the $Hon$, $Hirr$ and $Hp$ lines is evident and each line follows a $(1-T/T_c)^m$ dependence with $m$ = 1.8, 1.6, and 1.4 respectively. Figure 5 also suggests a collapse of $H^*$ values ( for $H$$\parallel$$ab$ and $H$45$^o$$ab$) which lie between $Hon$ and $Hp$. Values of $H_2$ (for $H$$\parallel$$c$) lie between  $Hp$ and $Hirr$. One may associate the anisotropic vortex dynamics observed here with the anisotropic neutron spin resonance found on a similar optimally doped sample\cite{luo2}, which is corroborated by isotropic vortex dynamics\cite{8} and isotropic neutron spin resonance\cite{luo3} found on overdoped samples of BaFe$_{2-x}$Ni$_x$As$_2$. Although we could not find evidence for a pinning crossover associated with $Hp$, we notice that the peak field position $Hp$ is time dependent, as predicted within the plastic pinning scenario\cite{abulafia}. For this reason, we try to fit the $Hp$ line for each geometry by the  proper expression\cite{abulafia}  $Hp$$\approx$$(1-(T/T_c)^4)^{1.4}$ which failed.  This fact concur with the experimental evidence that, $Hp$ in the studied system, is not related to a pinning crossover. 

Although our results do not show a direct evidence for a vortex lattice phase transition, neither the literature seems to present studies of the vortex structure for intermediate and high fields for the studied system, the strong evidence of absence of a pinning crossover associated with the peak effect motivated us to explore the possibility that the peak effect in the critical current might be due to a softening of the  vortex lattice associated with a phase transition as predicted in Ref.\onlinecite{rosenstein}.  The lower inset of Fig. 1b  shows a fitting of $J_c(H)$ at 14 K for $H$$\parallel$$c$ to the expression $J_c(B)$$=$$A/[(B-Bp)^2+(\Delta B)^2]^{5/4}$ of Ref.\onlinecite{rosenstein} where $A$ is a fitting parameter $Bp$ is the peak position and $\Delta B$ is the peak width. The resulting fitting conducted in a wide field range is excellent  and produced $Bp$ = 14 kOe and $\Delta B$ = 24 kOe. Similar fittings were obtained at different temperatures with $H$$\parallel$$c$.

In conclusion, our study of  flux-creep data based on the rate of magnetic relaxation, and on isofield activation energy curves, did not show any evidence of a pinning crossover occurring near the second magnetization peak of $M(H)$ curves in BaFe$_{1.9}$Ni$_{0.1}$As$_2$. Curves of $R~ vs.~ H$ for $H$$\parallel$$ab$ and $H$45$^o$$ab$-planes show a minimum at $H^*$ well below $Hp$, which behavior is quite similar to that found for an overdoped crystal of the same system for $H$$\parallel$$ab$ and $H$$\parallel$$c$ \cite{8}.  Curves of $R~ vs.~ H$ for $H$$\parallel$$c$ show a somewhat different behavior with a minimum at $H_1$ associated to $Hon$, and a maximum at $H_2$ well above $Hp$. A fitting based on the expression predicted by plastic pinning failed to explain the $Hp$ line of each phase diagram.

SSS, LG and ADA thank support from the Brazilian agencies CNPq and FAPERJ. LFC thanks the UK Funding Council the EPSRC grant EP/H040048. 

\end{document}